\documentclass[12pt,a4paper]{elsart}
\usepackage{latexsym,amssymb,epsfig}
\usepackage{amsmath,eepic}

\newcommand{\di}{\ensuremath{\mathrm{d}}}
\newcommand{\0}{$\phantom{0}$}
\newcommand{\DMS}{\mbox{\ensuremath{\mathbf{\scriptstyle^-}}\negthickspace\negthickspace {\sl D}}}
\renewcommand{\thetable}{\Roman{table}} 

\def\0{\phantom{0}}

\begin{document}

\pagenumbering{arabic} \baselineskip25pt
\begin{center}
{\bf \large
Self Diffusion and Binary Maxwell-Stefan Diffusion in Simple Fluids with the Green-Kubo Method} \\
\bigskip

G. A. Fern\'{a}ndez\footnotemark[1], J. Vrabec\footnotemark[1]$^,$\footnotemark[2], and
H. Hasse\footnotemark[1]
\end{center}
\footnotetext[1]{Institute of Thermodynamics and Thermal Process Engineering, University
of Stuttgart,\\ D-70550 Stuttgart, Germany} \footnotetext[2]{To whom correspondence
should be addressed, tel.: +49-711/685-6107, fax: +49-711/685-7657, email:
vrabec@itt.uni-stuttgart.de}
\renewcommand{\thefootnote}{\alph{footnote}}

\vskip3cm Number of pages: 26

Number of tables: 2

Number of figures: 5

Running title: Self and Binary Maxwell-Stefan Diffusion
\bigskip

\clearpage {\bf ABSTRACT}\\
Self diffusion coefficients and binary Maxwell-Stefan diffusion coefficients were
determined by equilibrium molecular dynamics simulations with the Green-Kubo method. The
study covers five pure fluids: neon, argon, krypton, xenon, and methane and three binary
mixtures: argon+krypton, argon+xenon, and krypton+xenon. The fluids are modeled by
spherical Lennard-Jones pair-potentials, with parameters which were determined solely on
the basis of vapor-liquid equilibria data. The predictions of the self diffusion
coefficients agree within $5 \%$ for gas state points and about $10 \%$ for liquid state
points. The Maxwell-Stefan diffusion coefficients are predicted within $10 \%$. A test
of Darken's model shows good agreement.

\clearpage 
{\bf KEYWORDS:} diffusion coefficients; Green-Kubo; Lennard-Jones; molecular
dynamics; molecular simulation; time correlation functions.

\clearpage

\textbf{1. INTRODUCTION}\\
Diffusion plays an important role in many chemical processes, such as catalysis or
adsorption. On the other hand, the measurement of diffusion coefficients is a time
consuming and difficult task \cite{kestin}. Molecular simulation offers the possibility
to straightforwardly determine diffusion coefficients on the basis of a given molecular
model. Both self diffusion coefficients and binary Maxwell-Stefan (MS) diffusion
coefficients can be obtained by non-equilibrium molecular dynamics (NEMD) or equilibrium
molecular dynamics (EMD). In this work, EMD is chosen.

From the pioneering work of Alder and Wainwright with hard spheres \cite{alder1,alder},
the simulation of diffusion coefficients has been an area of continuous research. There
are several contributions in which self diffusion coefficients
\cite{michels,rowley,meier}, binary \cite{jolly,schoen,heyes1,pas,zhou,lee} and ternary
diffusion coefficients \cite{vanl1,vanl2} for noble gases, methane and mixtures of these
are calculated. Less frequently, investigations with multi-center Lennard-Jones models,
e.g. mixtures of CH$_4$+SF$_6$ \cite{hoheisel} and CH$_4$+CF$_4$ \cite{schoen2}, or
polar fluids \cite{luo,vanl3} have been performed. With the exception of Refs. 5 and 6,
all investigations from the literature cover only diffusion coefficients in the liquid
phase and only for a limited range of state points.

This is the aim of the present work in which, as a first step, only simple fluids are
considered. Self diffusion coefficients for five pure fluids: neon, argon, krypton,
xenon, and methane (both liquid and gas) and three binary mixtures: argon+krypton,
argon+xenon, and krypton+xenon (gas) are predicted based on molecular models from the
literature and compared with experimental data. The pure component parameters of these
models were determined from vapor-liquid equilibria data alone \cite{vrabec2}. Binary
mixtures were modeled using one adjustable parameter for the unlike interaction which
was fitted to vapor-pressure data of the mixtures \cite{stoll1,vrabec3}. Throughout the
present work, for the molar mass the standard value from the literature is used
\cite{poling}. The simulation results on diffusion coefficients from the present work
are therefore predicted from vapor-liquid equilibria alone and obtained without any
adjustment to diffusion or other transport data. The studied systems are those for which
both molecular models and experimental data were available.

\clearpage
\textbf{2. METHOD}\\
2.1. Molecular Models

In this work,  only noble gases and methane are considered, so that the description of
the molecular interactions by the Lennard-Jones 12-6 (LJ) potential is sufficient and
physically meaningful. The LJ potential $u$ is defined by
\bigskip
\begin{equation}\label{lj126}
u_{ij}(r)=4 \epsilon_{ij} \left[\left(\frac{\sigma_{ij}} r
\right)^{12}-\left(\frac{\sigma_{ij}} r \right)^6\right],
\end{equation}

where $i$ and $j$ are the species indices, $\sigma_{ij}$ is the LJ size parameter,
$\epsilon_{ij}$ the LJ energy parameter, and $r$ the center-center distance between two
molecules. Pure substance parameters $\sigma_{ii}$ and $\epsilon_{ii}$ are taken from
Ref. 19 as given in Table \ref{T1}. They were adjusted by Vrabec et {\it al.}
\cite{vrabec2} to experimental pure substance vapor-liquid equilibrium data alone. For
modeling mixtures, parameters for the unlike interactions are needed. Following previous
work \cite{stoll1,vrabec3}, they are given by a modified Lorentz-Berthelot combination
rule

\bigskip
\begin{equation}\label{berte}
\sigma_{12}=\frac{(\sigma_{11}+\sigma_{22})}{2},
\end{equation}
and
\begin{equation}\label{lorenz}
\epsilon_{12}=\xi \cdot \sqrt{\epsilon_{11}\epsilon_{22}}.
\end{equation}
where $\xi$ is an adjustable binary interaction parameter. This parameter allows an
accurate description of the binary mixture data \cite{stoll1,vrabec3} and was determined
by an adjustment to one experimental vapor-liquid equilibrium state point. As for the
mixture krypton+xenon no binary interaction parameter is available, $\xi=$1 was assumed.
The binary interaction parameters are listed in Table \ref{T2}. \clearpage
2.2. Diffusion Coefficients

Diffusion coefficients can be calculated by equilibrium molecular dynamics through the
Green-Kubo formalism \cite{kubo,green}. In this formalism, transport coefficients are
related to integrals of time-correlation functions. There are various methods to relate
transport coefficients to time-correlation functions; a good review can be found in
\cite{zwanzig}. The self diffusion coefficient $D_i$ is given by \cite{vanl1}
\bigskip
\begin{equation}
\label{self} D_i=\frac{1}{3N_i}\int_{0}^{\infty} \di t \Big\langle\sum_{k=1}^{N_i}
\mathbf{v}_i^k(0) \cdot \mathbf{v}_i^k(t) \Big\rangle,
\end{equation}
where $\mathbf{v}_i^k(t)$ expresses the velocity vector of molecule $k$ of species $i$
and the notation $<...>$ denotes the ensemble average. Equation (\ref{self}) yields the
self diffusion coefficient for component $i$ averaging over $N_i$ molecules. The
expression for the binary Maxwell-Stefan diffusion coefficient $\DMS_{12}$ is given by
\cite{vanl1}
\bigskip
\begin{equation}
\label{mutual} \DMS_{12} = \frac{x_2}{3N_1} \left(\frac{x_1 M_1+x_2 M_2}{x_2M_2}
\right)^2 \int_{0}^{\infty} \di t~\Big\langle \sum_{k=1}^{N_1} \mathbf{v}_1^i(0) \cdot
\sum_{k=1}^{N_1} \mathbf{v}_1^i(t) \Big\rangle,
\end{equation}
where $M_i$ denotes the molar mass of molecules of species $i$, $N_1$ the number of
molecules of species 1 and $x_1$
and $x_2$ are the mole fractions.

To compare MS diffusion coefficients to available experimental data, it is necessary to
transform the MS diffusion coefficients to Fick diffusion coefficients. There is a
direct relation between binary MS diffusion coefficients $\DMS_{12}$, and binary Fick
diffusion coefficients $D_{12}$ \cite{krishna}, which is given by
\bigskip
\begin{equation}
\label{fick} D_{12}=\DMS_{12} \cdot Q,
\end{equation}
with
\bigskip
\begin{equation}
\label{factorq} Q =\frac{x_1}{k_{B}T}\left( \frac{\partial \mu_1}{\partial x_1}
\right)_{T,p},
\end{equation}

where $\mu_1$ is the chemical potential of species 1, $k_B$ the Boltzmann constant and
$T$ the temperature.
\bigskip

Because the present simulations provide both binary MS diffusion coefficients and self
diffusion coefficients, it is possible to test the often used model of Darken
\cite{darken,cussler}. It gives an estimate of the MS diffusion coefficient,
$\DMS_{12}^0$, from the self diffusion coefficients of both components in a binary
mixture $D_{1}$ and $D_{2}$,
\bigskip
\begin{equation}
\label{ideal} \DMS_{12}^0= D_{1} \cdot x_1+D_{2} \cdot x_2.
\end{equation}

\bigskip
2.3. Simulation Details

The molecular simulations were performed in a cubic box of volume $V$ containing
standard $N=500$ molecules modeled by the LJ potential. The cut-off radius was set to
$r_{c}=5 \sigma $; standard techniques for periodic boundary conditions and long-range
corrections were used \cite{allen}. The simulations were started with the molecules in a
face-center-cubic lattice with random velocities, the total momentum of the system was
set to zero, and Newton's equations of motion were solved with a velocity-Verlet
algorithm \cite{allen}. The time step for this algorithm was set to $\Delta t \cdot
\sqrt{\epsilon_1 / m_1}/\sigma_1=0.001$ for liquid and $0.01$ for gas state points. The
diffusion coefficients were calculated in a NVE ensemble, using Eqs. (\ref{self}) and
(\ref{mutual}). The relative fluctuation in the total energy in the NVE ensemble was
less than $10^{-3}$ for the longest run, which yields a temperature drift of less than
$0.5$ K. The simulations are initiated in a NVT ensemble until equilibrium at the
desired density and temperature is reached. $25~000$ time steps were used for that
equilibration. Once the equilibrium is reached, the thermostat is turned off, and then
the NVE ensemble is invoked. The experimental data which was used for comparison to our
simulations is often reported at given pressure and temperature. In these cases, a prior
isobaric-isothermal NpT simulation \cite{andersen} was performed, from which the
corresponding densities for the NVE ensemble were taken. The statistical uncertainty of
the diffusion
coefficients was estimated with the standard block average technique \cite{frenkel}.

\begin{figure}[ht]
\caption[Large plot: Velocity autocorrelation functions. Small plot: Integral following
Eq. (\ref{self}). Both plots are shown for selected state points of argon: {------}
$T$=77.7 K and $\rho$=163.26 mol $\cdot$ m$^{-3}$; {-- -- -- --} $T$=353.2 K and
$\rho$=34.49 mol $\cdot$ m$^{-3}$.]{} \label{fig1y3}
\begin{center}
\includegraphics[width=150mm,height=200mm]{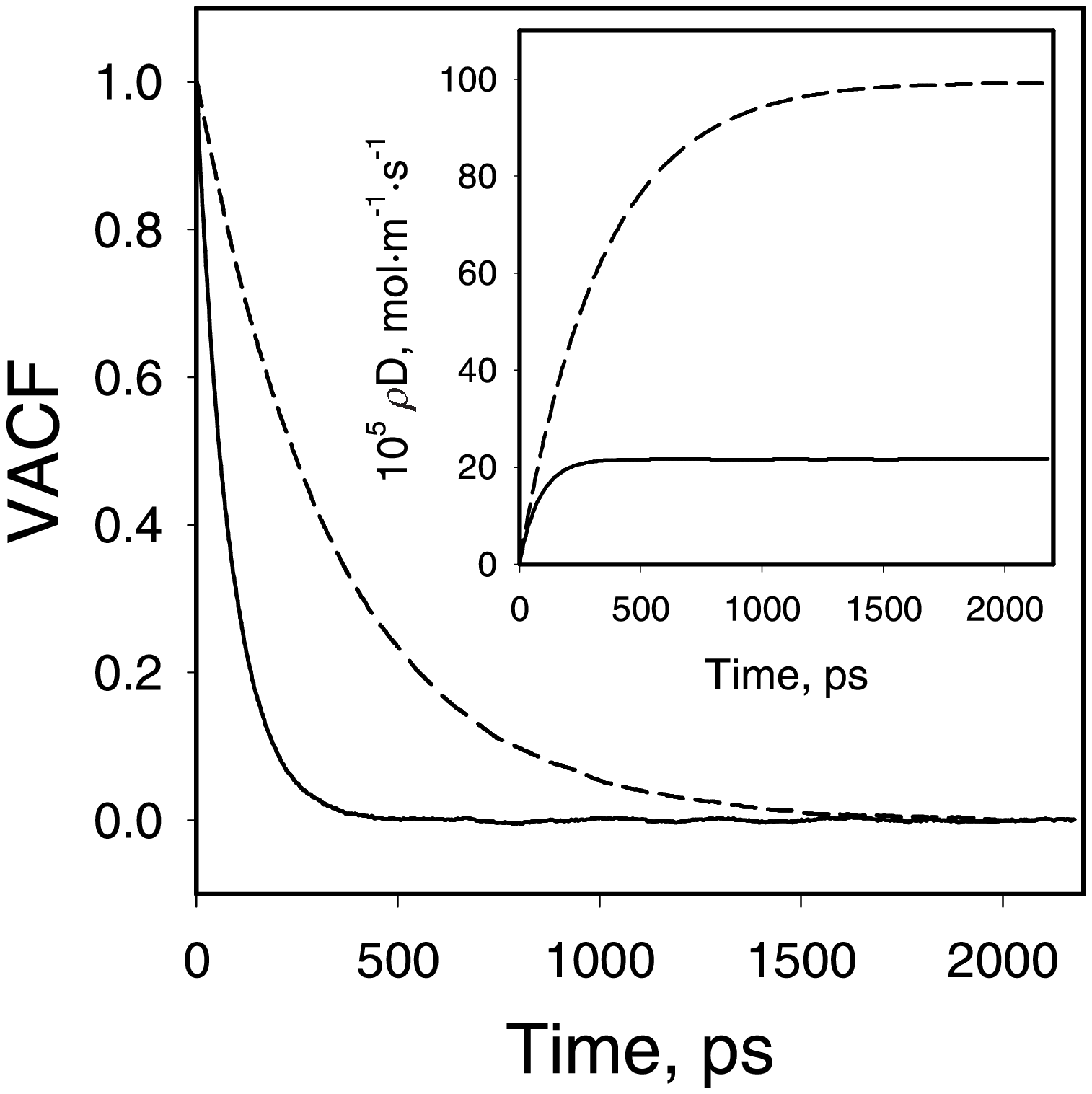}
\end{center}
\end{figure}

The self diffusion coefficient is a property related to one molecule. It is possible to
obtain very good statistics with a few independent velocity autocorrelation functions
(VACF). The self diffusion coefficients were calculated by averaging over $200$
independent VACF each with $500$ molecules, i.e. a total of $10^5$ VACF. For gas
densities, the VACF decays very slowly and therefore long simulation runs were necessary
in order to achieve the VACF decay and hence independent time-origins. Here, a
compromise between simulation time and time-origin independence had to be made. In order
to keep the simulation time low, and following the work of Schoen and Hoheisel
\cite{schoen}, the separation between time origins was chosen at least as long as the
VACF needs to decay to $1/e$ of its normalized value. The choice of this separation time
and the production phase depended upon the temperature and density conditions. In
theory, as Eq. (\ref{self}) shows, the value of the diffusion coefficient is determined
by an infinite time integral. In fact, however, the integral is evaluated based on the
length of the simulation. The integration must be stopped at some finite time, ensuring
that the contribution of the long-time tail \cite{alder} is small compared to the
desired statistical uncertainty of the diffusion coefficient. Figure \ref{fig1y3} shows
the behavior of the VACF and its integral given by Eq. (\ref{self}) for two selected gas
state points of argon. As can be seen, for the higher density state point, the VACF has
decayed after $500$ ps to less than $1 \%$ of its normalized value. Later it oscillates
around zero. The same can be seen after $1500$ ps at the lower density state point. It
was assumed here that the VACF has fully decayed when these oscillations reached a
threshold below $0.5 \%$ of their normalized value. Furthermore, a graphical inspection
of the VACF integral was made, in order to verify a
sufficient integration time.

An important time scale to calculate the VACF is the time that a sound wave takes to
cross the simulation box. VACF calculated for times higher than that may show a
systematic error \cite{hansen,heyes2}. That criterion was verified using the
experimental speed of sound taken from \cite{cook}. For the simulations of gases, the
VACF decay time was found to be higher than that time. To check whether this to leads to
a systematic error in the present simulations, the system size was varied. For the
lowest density state point of argon, where the above mentioned problem would be expected
to be most severe, simulations with a constant number of time origins and increasing
system sizes were carried out. System sizes of $N=864$, $2048$, $4000$, and $6912$
molecules were investigated. All results were found to agree within the statistical
uncertainty, and no size dependence could be observed. It is therefore concluded
that no systematic error due to system size in gas phase simulations is present.

The Maxwell-Stefan diffusion coefficient is a collective quantity and therefore the
statistics can only be improved by averaging over longer simulation runs. The MS
diffusion coefficients were calculated by averaging over $2000$ velocity correlation
functions (VCF) as proposed by Schoen and Hoheisel \cite{schoen}. In order to obtain
independent time origins, similar criteria as employed for the self diffusion
coefficients were used to determine the necessary length of the VCF.

\clearpage
\textbf{3. RESULTS}

3.1. Self Diffusion Coefficients

\begin{figure}[ht]
\caption[Self diffusion coefficients of neon, argon and krypton (gas phase) predicted by
molecular simulation compared to experimental data \cite{cook} at $p$=0.1013 MPa. neon:
{$\square$} exp., {$\blacksquare$} sim.; argon: {$\vartriangle$} exp.,
{$\blacktriangle$} sim.; krypton: {\large $\circ$} exp., {\large $\bullet$} sim. The
solid lines represent the results of the correlation of Liu et {\it al}. \cite{liu}.
]{}\label{fig2}
\begin{center}
\includegraphics[width=150mm,height=200mm]{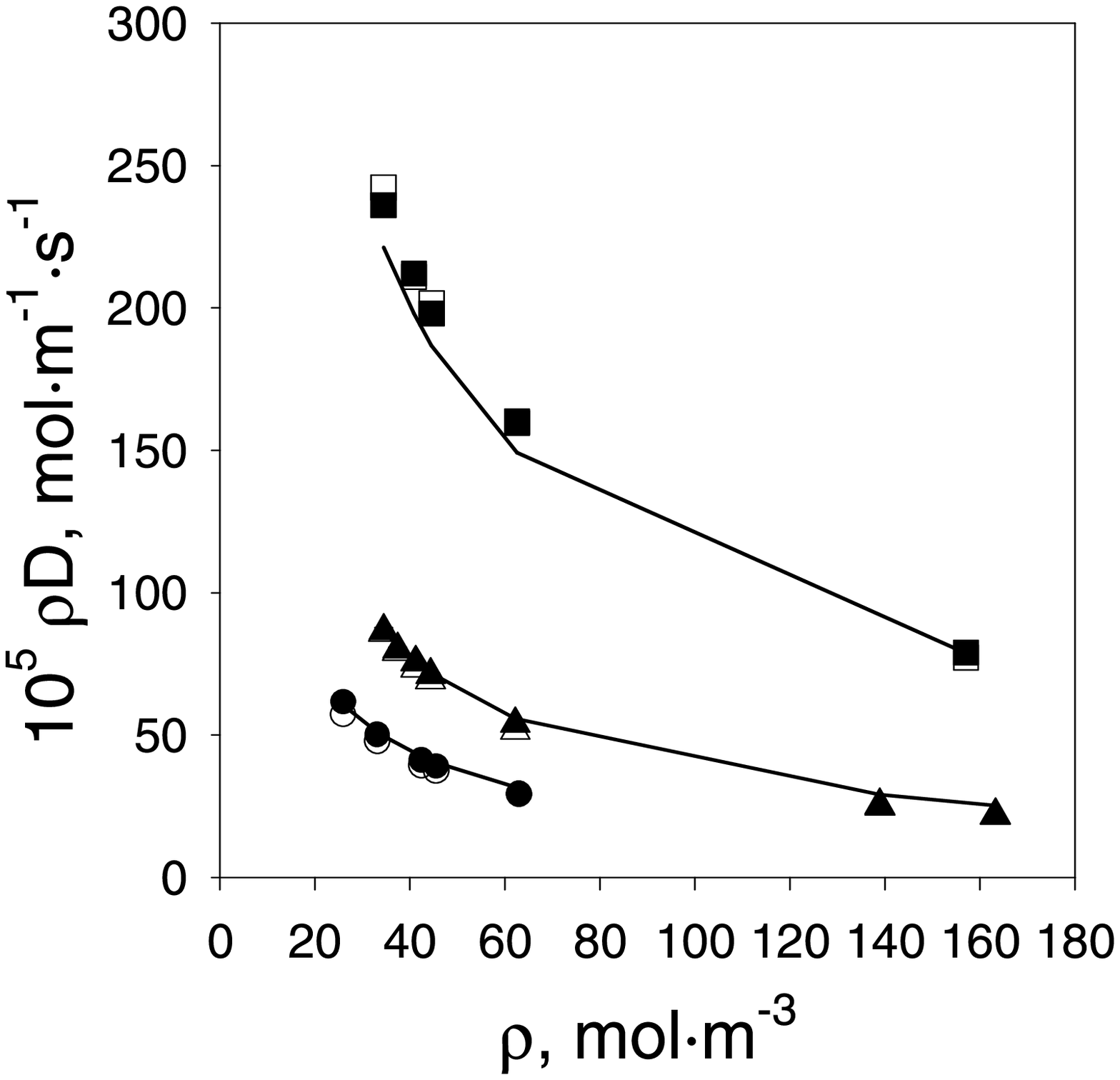}
\end{center}
\end{figure}

\begin{figure}[ht]
\caption[Self diffusion coefficients of neon, argon, krypton, xenon, and methane (liquid
and gas phase) predicted by molecular simulation compared to experimental data
\cite{bewilogua,mifflin,codastefano,peereboom,harris} at constant temperatures and
different pressures. neon $T$=37 K: {$\square$} exp., {$\blacksquare$} sim.; argon
$T$=323 K: {$\vartriangle$} exp., {$\blacktriangle$} sim.;
 krypton $T$=273 K: {\large $\circ$} exp., {\large $\bullet$} sim.; xenon $T$=298 K: {\large $\lozenge$} exp., {\large $\blacklozenge$} sim.;
 methane $T$=298 K: {$\triangledown$} exp., {$\blacktriangledown$} sim. The
solid lines represent the results of the correlation of Liu et {\it al}. \cite{liu}.]{}
\label{fig3}
\begin{center}
\includegraphics[width=150mm,height=200mm]{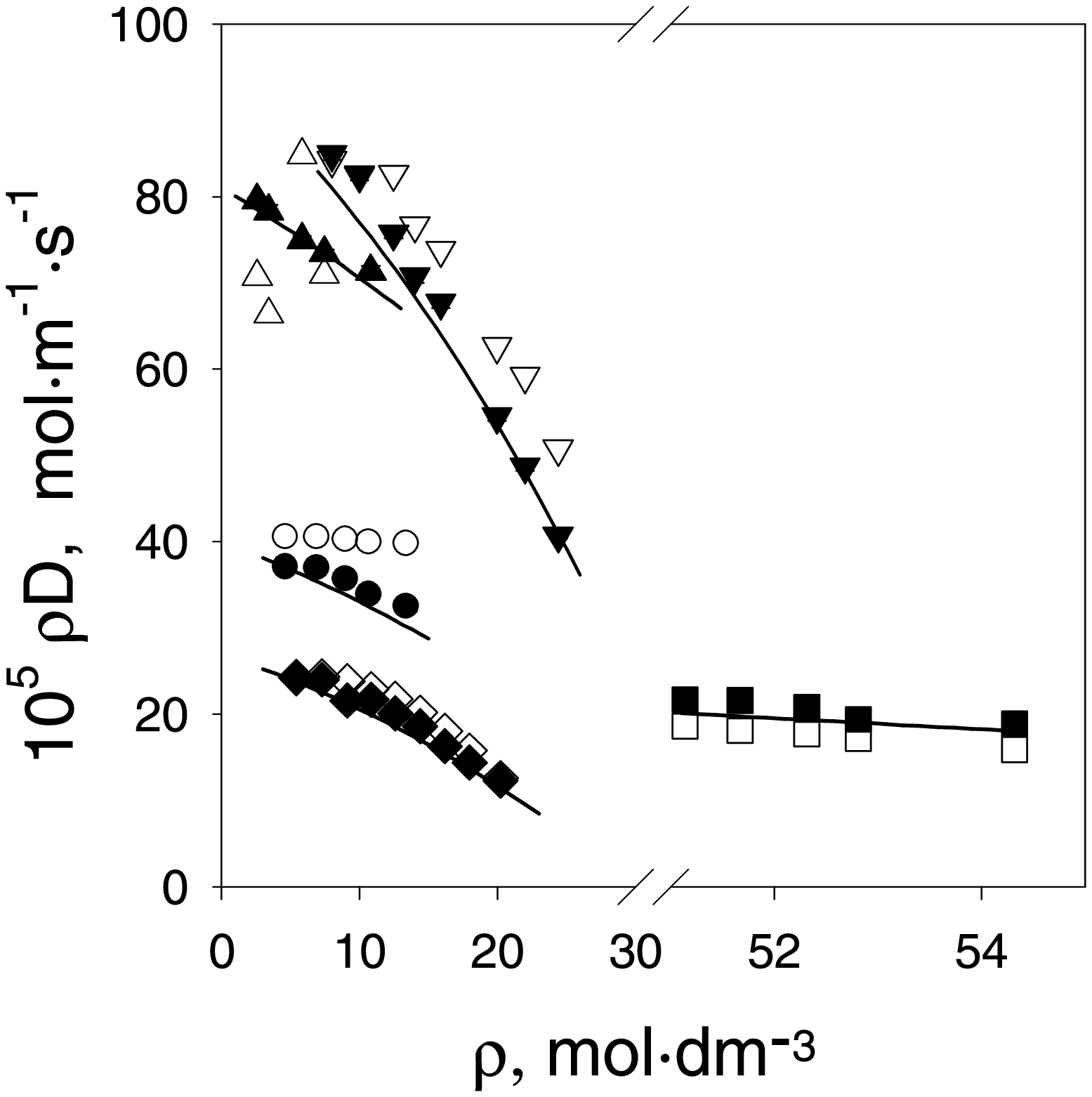}
\end{center}
\end{figure}

Figure \ref{fig2} shows the results for the self diffusion coefficients of neon, argon,
and krypton compared with experimental data for gas state points \cite{cook}. The lines
in Fig. \ref{fig2} are the results of the correlation of Liu et {\it al}. \cite{liu}
using the LJ parameters from Table \ref{T1}. The data are given at constant pressure and
at different temperatures. Figure \ref{fig3} shows the results for the self diffusion
coefficients for neon, argon, krypton, xenon, and methane compared with experimental
data \cite{bewilogua,mifflin,codastefano,peereboom,harris} for liquid state points. The
lines in Fig. \ref{fig3} are the results of the correlation of Liu et {\it al}.
\cite{liu} using the LJ parameters from Table \ref{T1}. The data are given at constant
temperature and at different pressures. Overall, a very good agreement between the
predictions and the experimental data is found. The best results are found for neon,
argon, and krypton in the gas phase with deviations within a few percent. The results
for liquid state points show somewhat higher relative deviations from the experimental
data (around 10$\%$). It can be seen that the correlation agrees reasonably well with
the simulation data, typical deviations are about $5 \% $. This accuracy lies in the
range claimed by the authors of Ref. 35.

3.2. Binary Maxwell-Stefan Diffusion Coefficients

Binary MS diffusion coefficients were calculated for the gaseous mixtures argon+krypton,
argon+xenon, and krypton+xenon. The results are compared to experimental Fick diffusion
coefficients. The thermodynamic factor $Q$, that relates the MS diffusion coefficient to
the Fick diffusion coefficient, cf. Eq. (\ref{fick}), is assumed to be unity for all
cases studied here. This is supported by the calculations of several authors
\cite{jolly,schoen,zhou,mills}. As a test, $Q$ was estimated by two simulations to
calculate a finite difference \cite{carnahan} for each mixture at the most dense state
point. The assumption $Q=$1 was confirmed within the
statistical uncertainty of the calculations.

\begin{figure}[ht]
\caption[Binary Maxwell-Stefan diffusion coefficients for gaseous equimolar mixtures of
argon+krypton predicted by molecular simulation compared to experimental data
\cite{shankland} at $T$=323.16 K : argon+krypton {$\triangle$} exp., {$\blacktriangle$}
sim. The solid lines represent the results of the correlation of Darken
\cite{darken,cussler}.]{} \label{fig4}
\begin{center}
\includegraphics[width=150mm,height=200mm]{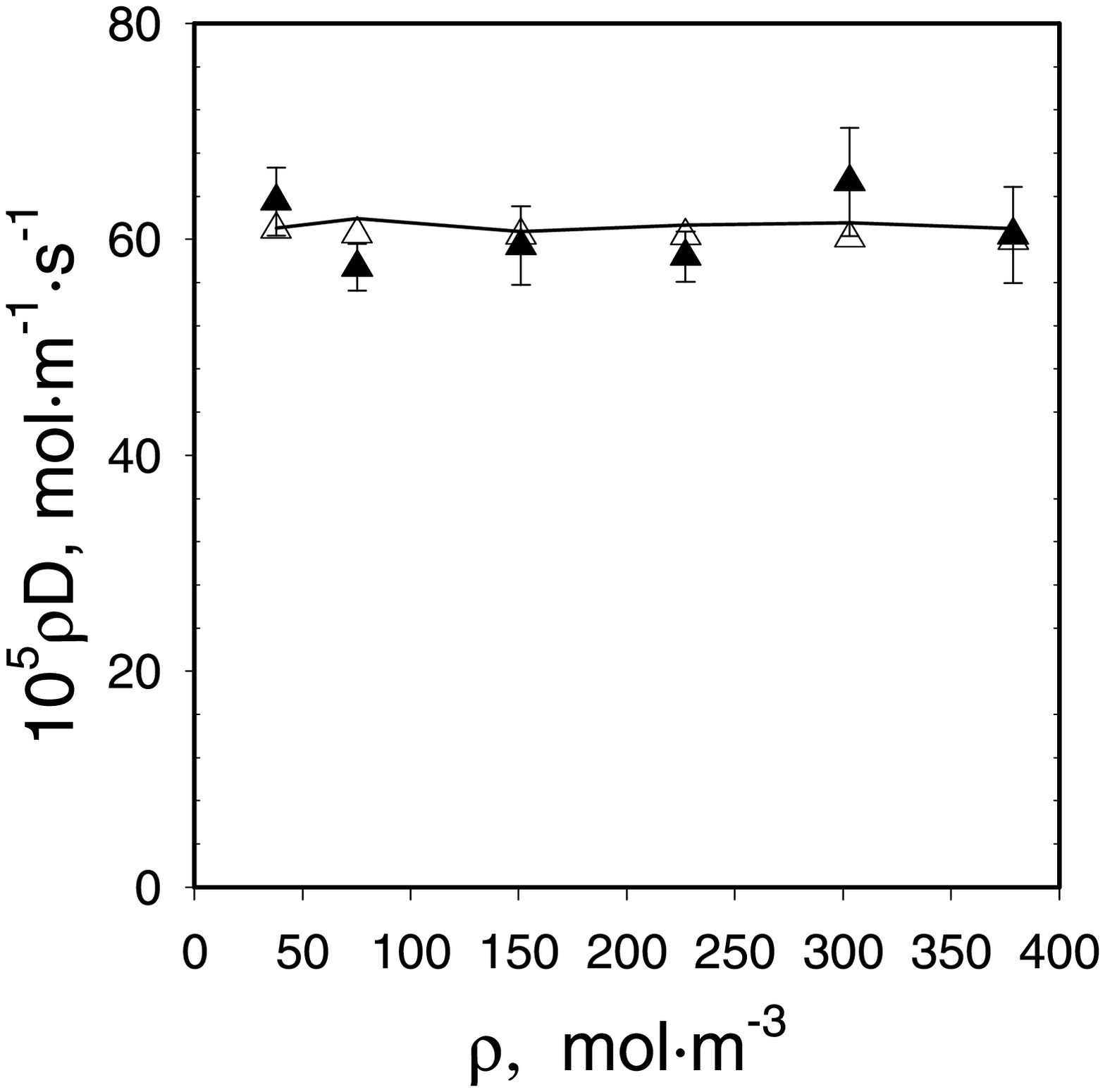}
\end{center}
\end{figure}

\begin{figure}[ht]
\caption[Binary Maxwell-Stefan diffusion coefficients for gaseous equimolar mixtures of
argon+xenon and krypton+xenon predicted by molecular simulation compared to experimental
data at $p$=0.1013 MPa: argon+xenon: {\large $\circ$} exp., {\large $\bullet$} sim.;
krypton+xenon {$\triangledown$} exp., {$\blacktriangledown$} sim. The solid lines
represent the results of the correlation of Darken \cite{darken,cussler}.]{} \label{fig5}
\begin{center}
\includegraphics[width=150mm,height=200mm]{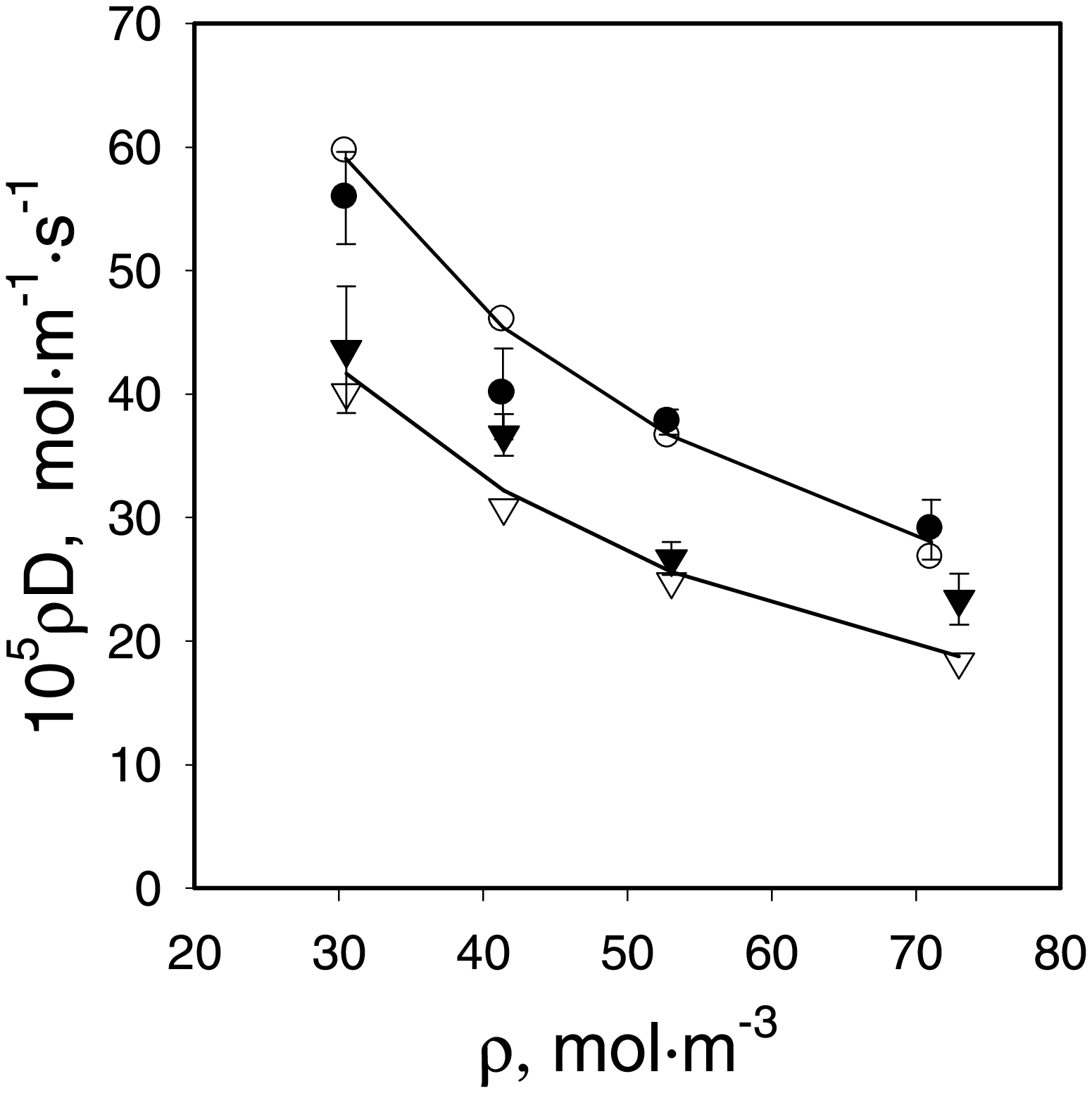}
\end{center}
\end{figure}

Figure \ref{fig4} shows the simulation results for the mixture argon+krypton in
comparison to experimental data taken from Ref. 43. The continuous line in Fig.
\ref{fig4} are the results of the correlation of Darken \cite{darken,cussler}. In this
case, the experimental data \cite{vanh} were reported at constant temperature. Figure \ref{fig5}
shows the results for the mixtures argon+xenon and krypton+xenon at constant pressure.
Good agreement between the predictions and the experimental values is found.
The best results are observed for the mixture argon+krypton. This mixture shows typical
relative deviations of $4 \%$ from the experimental data; the corresponding numbers are
$8 \%$ for argon+xenon, and $16 \%$ for krypton+xenon. It must be pointed out that for
the mixture krypton+xenon, no binary interaction parameter $\xi$ was available.

In Figs. \ref{fig4} and \ref{fig5}, it is interesting to analyze the performance of the
empirical model of Darken for estimating MS diffusion coefficients in the gas phase on
the basis of known binary self diffusion coefficients. The figures show that the model
of Darken agrees very well with the binary MS diffusion coefficients, typical deviations
are about $5 \%$.

\clearpage
\textbf{4. CONCLUSION}

In the present work, molecular models of simple fluids that were adjusted to
vapor-liquid equilibrium only were used to predict self and MS diffusion coefficients.
The diffusion coefficients were determined with molecular dynamics simulations using the
Green-Kubo method. Five pure fluids and three binary mixtures were studied covering a
broad range of state points. The fluids were modeled with the Lennard-Jones pair
potential with parameters taken from the literature. It is found that the prediction of
diffusion coefficients from vapor-liquid equilibrium data using that simple model yields
good results. This supports the finding that the spherical LJ 12-6 potential is an
adequate description for the regarded noble gases and methane. When molecular models are
adjusted to diffusion coefficient data, excellent descriptions can be expected. It is
worthwhile to extend the study to more complex fluids.

\clearpage

\textbf{ACKNOWLEDGMENT}

The authors thank Jochen Winkelmann, University of Halle-Wittenberg, for the diffusion
coefficient data and J\"urgen Stoll, University of Stuttgart, for fruitful discussions.

\clearpage
\textbf{REFERENCES}

$\0$1. J. Kesting and W.A. Wakeham, {\it Transport Properties of Fluids: Thermal\\
$\0\0\0$Conductivity, Viscosity and Diffusion Coefficient}, CINDAS Data Series in \\
$\0\0\0$Material Properties (Hemisphere Publishing, New York, 1988).

$\0$2. B.J. Alder and T.E. Wainwright,{\it Phys. Rev. A.} {\bf 18}:988 (1967).

$\0$3. B.J. Alder and T.E. Wainwright, {\it Phys. Rev. A.} {\bf 1}:18 (1970).

$\0$4. J.P.J. Michels and N.J. Trappeniers, {\it Physica A.} {\bf 90A}:179 (1978).

$\0$5. R.L. Rowley and M.M. Paiter, {\it Int. J. Thermophys.} {\bf 18}:1109 (1997).

$\0$6. K. Meier, A. Laesecke, and S. Kabelac, {\it Int. J. Thermophys.} {\bf 22}:161
(2001).

$\0$7. D. Jolly and R. Bearman, {\it Mol. Phys.} {\bf 41}:137 (1980).

$\0$8. M. Schoen and C. Hoheisel, {\it Mol. Phys.} {\bf 52}:33 (1984).

$\0$9. D.M. Heyes and S.R. Preston, {\it Phys. Chem. Liq.} {\bf 23}:123 (1991).

10. M.F. Pasand and B. Zwolinski, {\it Mol. Phys.} {\bf 73}:483 (1991).

11. Y. Zhou and G.H. Miller, {\it Phys. Rev. E.} {\bf 53}:1587 (1995).

12. J.C. Lee, {\it Physica A.} {\bf 247}:140 (1997).

13. I.M.J.J. van de Ven-Lucassen, T.J.H. Vlugt, A.J.J. van der Zenden,
 and P.J.A.M.\\
$\0\0\0$Kerkhof, {\it Mol. Phys.} {\bf 94}:495 (1998).

14. I.M.J.J. van de Ven-Lucassen, A.M.V.J. Otten, T.J.H. Vlugt, and P.J.A.M. Kerkhof,\\
$\0\0\0${\it Mol. Simulation} {\bf 23}:495 (1998).

15. C. Hoheisel, {\it J. Chem. Phys.} {\bf 89}:3195 (1988).

16. M. Schoen and C. Hoheisel, {\it J. Chem. Phys.} {\bf 58}:699 (1986).

17. H. Luo and C. Hoheisel, {\it Phys. Rev. A.} {\bf 43}:1819 (1991).

18. I.M.J.J. van de Ven-Lucassen, T.J.H. Vlugt, A.J.J. van der Zanden, and P.J.A.M.\\
$\0\0\0$ Kerkhof, {\it Mol. Simulation} {\bf 23}:79 (1999).

19. J. Vrabec, J. Stoll, and H. Hasse, {\it J. Phys. Chem. B.} {\bf 105}:12126 (2001).

20. J. Stoll, J. Vrabec, and H. Hasse, {\it AIChE J.} {\bf 49}:2187 (2003).

21. J. Vrabec, J. Stoll, and H.Hasse, {\it Molecular Model of unlike Interactions in
Mixtures},\\
$\0\0\0$ Institute of Thermodynamics and Thermal Process Engineering,University of\\
$\0\0\0$ Stuttgart (2003).

22. B.E. Poling, J.M. Prausnitz, and J.P. O'Connell, {\it The Properties of Gases and
Liquids},\\
$\0\0\0$5th Edition (McGraw-Hill, New York, 2001).

23. R. Kubo, {\it J. Phys. Soc. Japan} {\bf 12}:570 (1957).

24. M.S. Green, {\it J. Chem. Phys.} {\bf 22}:398 (1954).

25. R. Zwanzig, {\it Ann. Rev. Phys. Chem.} {\bf 16}:67 (1965).

26. R. Taylor and R. Krishna, {\it Multicomponent Mass Transfer} (John Wiley {\&}
Sons,\\
$\0\0\0$New York, 1993).

27. L.S. Darken, {\it AIME} {\bf 175}:184 (1948).

28. E.L. Cussler, {\it Diffusion Mass Transfer in Fluid Systems} (Cambridge University\\
$\0\0\0$Press, Cambridge, 1997).

29. M.P. Allen and D.J. Tildesley, {\it Computer Simulation of Liquids} (Clarendon
Press,\\
$\0\0\0$Oxford, 1987).

30. H.C. Andersen, {\it J. Chem. Phys.} {\bf 72}:2384 (1980).

31. D. Frenkel and R. Smit, {\it Understanding Molecular Simulation} (Academic Press,
San\\
$\0\0\0$Diego, 1996).

32. J.P. Hansen and I.A. Mcdonald, {\it Theory of Simple Liquids} (Academic Press,
London,\\
$\0\0\0$1986).

33. D.M. Heyes, {\it The Liquid State: Applications of molecular Simulation} (John Wiley
{\&}\\
$\0\0\0$Sons, New York, 1998).

34. G.A. Cook, {\it Argon, Helium and the Rare Gases, Vol. I} (Interscience Publishers,
New\\
$\0\0\0$York, 1961).

35. H. Liu, C.M. Silva, and E.A. Macedo, {\it Chem. Eng. Sci.} {\bf 53}:2403 (1998).

36. L. Bewilogua, C. Gladun, and B. Kubsch, {\it J. Low Temp. Phys.} {\bf 92A}:315
(1971).

37. T.R. Mifflin and C.O. Bennett, {\it J. Chem. Phys.} {\bf 29}:975 (1958).

38. P. Codastefano, M.A. Ricci, and V. Zanza, {\it Physica A} {\bf 92A}:315 (1978).

39. P.W.E. Peereboom, H. Luigjes, and K.O. Prins, {\it Physica A} {\bf 156A}:260 (1989).

40. K.R. Harris, {\it Physica A} {\bf 94A}:448 (1978).

41. R. Mills, R.Malhotra, L.A. Woolf, and D.G. Miller, {\it J. Phys. Chem.} {\bf
98}:5565 (1994).

42. B. Carnahan, H.A. Luther, and J.O. Wilkes, {\it Applied Numerical Methods} (John
Wiley\\
$\0\0\0${\&} Sons, New York, 1969).

43. I.R. Shankland and P.J. Dunlop, {\it Physica A} {\bf 100A}:64 (1980).

44. R.J.J. van Heijningen, J.P. Harpe, and J.J.M. Beenakker, {\it Physica A} {\bf 38A}:1
(1968).

\clearpage

\begin{table}[t]
\noindent \caption{Potential Model Parameters for the Pure Fluids used in this
Work$^{a,b}$.}\label{T1}
\bigskip
\begin{center}
\begin{tabular}{|l||c|c|c|c|c|} \hline
Fluid    &  \0\0$\sigma$  (\r{A})   &  $ \epsilon /k_{\rm B}$ (K)  & $M$  (g $\cdot$
mol$^{-1}$)
\\\hline\hline
 $\rm  Neon $    & 2.8010         & \033.921       &  \020.180                                   \\ \hline
 $\rm  Argon $   & 3.3952         &  116.79\0      &  \039.948                                   \\ \hline
 $\rm  Krypton $ & 3.6274         &  162.58\0      &  \083.8\0\0                               \\ \hline
 $\rm  Xenon   $ & 3.9011         &  227.55\0      &   131.29\0                                   \\ \hline
 $\rm  Methane $ & 3.7281         &  148.55\0      &  \016.043                                    \\ \hline
\end{tabular}
\end{center}
\end{table}
\footnotetext[1]{Values taken from Ref. 19.} \footnotetext[2]{The Molar Mass $M$ was
taken from Ref. 22.} \clearpage\clearpage

\begin{table}[t]
\noindent \caption{Binary Interaction Parameters for the Binary Mixtures taken from Ref.
21.} \label{T2}
\bigskip
\begin{center}
\begin{tabular}{|l|c|} \hline
\0 Mixture \0     & \0\0 $\xi$ \0\0  \\ \hline\hline
 $\rm Argon+ \rm Krypton$     & 0.988    \\ \hline
 $\rm Argon+ \rm Xenon$     & 1.000    \\ \hline
 $\rm Krypton+ \rm Xenon$     & 0.989    \\ \hline
\end{tabular}
\end{center}
\end{table}
\clearpage \clearpage
\listoffigures \clearpage


\clearpage

\end{document}